\documentclass[a4paper,twocolumn,english,aps,prl,superscriptaddress]{revtex4-1}
\usepackage[T1]{fontenc}
\usepackage[latin9]{inputenc}
\usepackage{color}
\usepackage{babel}
\usepackage{amsmath}
\usepackage{amssymb}
\usepackage{graphicx}
\usepackage{esint}
\usepackage[unicode=true,pdfusetitle,
 bookmarks=true,bookmarksnumbered=false,bookmarksopen=false,
 breaklinks=false,pdfborder={0 0 1},backref=false,colorlinks=false]
 {hyperref}

\makeatletter

\usepackage{babel}

\makeatother

\begin{document}
\renewcommand{\figurename}{FIG.} 
\title{Klein-bottle quadrupole insulators and Dirac semimetals }
\author{Chang-An Li}
\email{changan.li@uni-wuerzburg.de}

\affiliation{Institute for Theoretical Physics and Astrophysics, University of
Würzburg, 97074 Würzburg, Germany}
\author{Junsong Sun}
\affiliation{School of Physics, Beihang University, 100191 Beijing, China}
\author{Song-Bo Zhang}
\affiliation{International Center for Quantum Design of Functional Materials (ICQD),
Hefei National Research Center for Physical Sciences at the Microscale,
University of Science and Technology of China, Hefei, Anhui 230026,
China}
\author{Huaiming Guo}
\affiliation{School of Physics, Beihang University, 100191 Beijing, China}
\author{Björn Trauzettel}
\affiliation{Institute for Theoretical Physics and Astrophysics, University of
Würzburg, 97074 Würzburg, Germany}
\affiliation{Würzburg-Dresden Cluster of Excellence ct.qmat, 97074 Würzburg, Germany}
\date{\today }
\begin{abstract}
The Benalcazar-Bernevig-Hughes (BBH) quadrupole insulator model is
a cornerstone model for higher-order topological phases. It requires
$\pi$ flux threading through each plaquette of the two-dimensional
Su-Schrieffer-Heeger model. Recent studies showed that particular
$\pi$-flux patterns can modify the fundamental domain of momentum
space from the shape of a torus to a Klein bottle with emerging topological
phases. By designing different $\pi$-flux patterns, we propose two
types of Klein-bottle BBH models. These models show rich topological
phases, including Klein-bottle quadrupole insulators and Dirac semimetals.
The phase with nontrivial Klein-bottle topology shows twined edge
modes at open boundaries. These edge modes can further support second-order
topology, yielding a quadrupole insulator. Remarkably, both models
are robust against flux perturbations. Moreover, we show that different
$\pi$-flux patterns dramatically affect the phase diagram of the
Klein-bottle BBH models. Going beyond the original BBH model, Dirac
semimetal phases emerge in Klein-bottle BBH models featured by the
coexistence of twined edge modes and bulk Dirac points. 
\end{abstract}
\maketitle

\section{Introduction}

The quadrupole insulator model proposed by Benalcazar, Bernevig, and
Hughes (BBH) \cite{Benalcazar17Science,BBH17prb} is an important
model for the study of higher-order topological insulators \cite{Langbehn17prl,SongZD17prl,PengY17prb,Khalaf18prb,Geier18prb,Schindler18NP,Serra-Garcia18nature,Peterson18nature,Franca18prb,Schindler18SA,Ezawa18prl,Miert18prb,Imhof18np,WangZJ19prl,Roy19prb,Roy19prr,XieBY19prl,ParkMJ19prl,Benalcazar19prb,Okugawa19prb,XLSheng19prl,ChenR20prl,Ghosh20prb,RenY20prl,ZhangRX20prl,Hirosawa20prl,LiCA20prl,QiY20prl,Choi20nm,LiC20prb,LiCA21prl,YangYB21prb,WeiQ21prl,LiuB21prl,Schulz22nc,Benalcazar22prl,Leiz22prb,ZhuY23prb,Luoxj23prb}.
It exhibits a quadrupole insulator phase with quantized bulk multipole
moments. This phase is characterized by corner states carrying fractional
corner charges $\pm e/2$, generalizing the bulk-boundary correspondence
to higher order. The basic construction unit of quadrupole insulators
is the Su-Schrieffer-Heeger (SSH) model \cite{SSH79prl}. The SSH
model possesses quantized dipole moments in the bulk. Since a quadrupole
consists of two separated dipoles, one may couple the SSH models in
a particular way to obtain quantized quadrupole moments in the two-dimensional
(2D) bulk. However, direct coupling of one-dimensional (1D) SSH chains
from two directions does not work. It results in a gapless 2D SSH
model with topological properties \cite{LiuF17prl,LiCA22prr}. To
get an insulating phase with quantized quadrupole moments, the indispensable
ingredients are $\pi$ fluxes threading each plaquette of the entire
2D lattice. The $\pi$-flux pattern generates an insulating phase
at half-filling. It projectively modifies the mirror symmetry $M_{x}$
and $M_{y}$ from commuting $[M_{x},M_{y}]=0$ to anticommuting $\{M_{x},M_{y}\}=0$.
This change results in the BBH model with quantized bulk quadrupole
moments \cite{Benalcazar17Science,BBH17prb}. 

In our work, we vary the $\pi$-flux pattern of the BBH model. This
modification gives rise to a new class of models that we call Klein-bottle
BBH models. The name stems from the shape of the fundamental domain
of momentum space being modified from a torus to a Klein bottle in
mathematics by particular $\pi$-flux patterns \cite{ChenZ22nc,ZhangC23prl,ChenZ23nc,ZhaoYX20prb,HuJ23arxiv,TaoY23arxiv2,XueH22prl,LiT22prl,BaoS23pra,JiangC23OL,WangY2023arxiv,ZhuZ23arxiv,TaoY23arxiv}.
In the first type of Klein-bottle BBH model, the $\pi$ fluxes are
applied only at the even numbered columns of plaquettes in the 2D
SSH lattice {[}see Fig.~\ref{fig:model I}(a){]}. This model supports
nontrivial Klein-bottle quadrupole insulator phases with corresponding
boundary signatures such as quantized edge polarizations and fractional
corner charges. We show that the nontrivial Klein-bottle quadrupole
insulator is robust against flux perturbations. In the second Klein-bottle
BBH model, we instead apply $\pi$ fluxes at the odd number of columns
of plaquettes {[}see Fig.~\ref{fig:model2}(a) below{]}. This subtle
difference of $\pi$-flux patterns dramatically changes the phase
diagram of the system. The second model does not support nontrivial
Klein-bottle quadrupole insulators anymore. It shares some features
with the first model, such as the twined edge modes and corner-localized
charges, but its insulator phase is trivial with vanishing bulk quadrupole
moments. Interestingly, we identify emergent Klein-bottle Dirac semimetal
phases in both models, characterized by the coexistence of twined
edge modes and bulk Dirac points. In particular, four Dirac points
are located at high symmetry points of the Brillouin zone (BZ). They
are related by glide-mirror symmetry in momentum space. There are
no such Dirac semimetal phases in the original BBH model. 

The article is organized as follows. In Sec. II, we present the first
Klein-bottle BBH model with an emphasis on the nontrivial Klein-bottle
quadrupole insulator phase. In Sec. III, we study Klein-bottle Dirac
semimetal phases. In Sec. IV, we show the robustness of Klein-bottle
quadrupole insulators against flux variations. In Sec. V, we consider
the properties of the second Klein-bottle BBH model with a different
$\pi$-flux pattern. Finally, we conclude our results with a discussion
in Sec.VI.

\section{Klein-bottle quadrupole insulators induced by $\mathbb{Z}_{2}$ gauge
fields}

\subsection{The first Klein-bottle BBH model}

\begin{figure}
\includegraphics[width=1\linewidth]{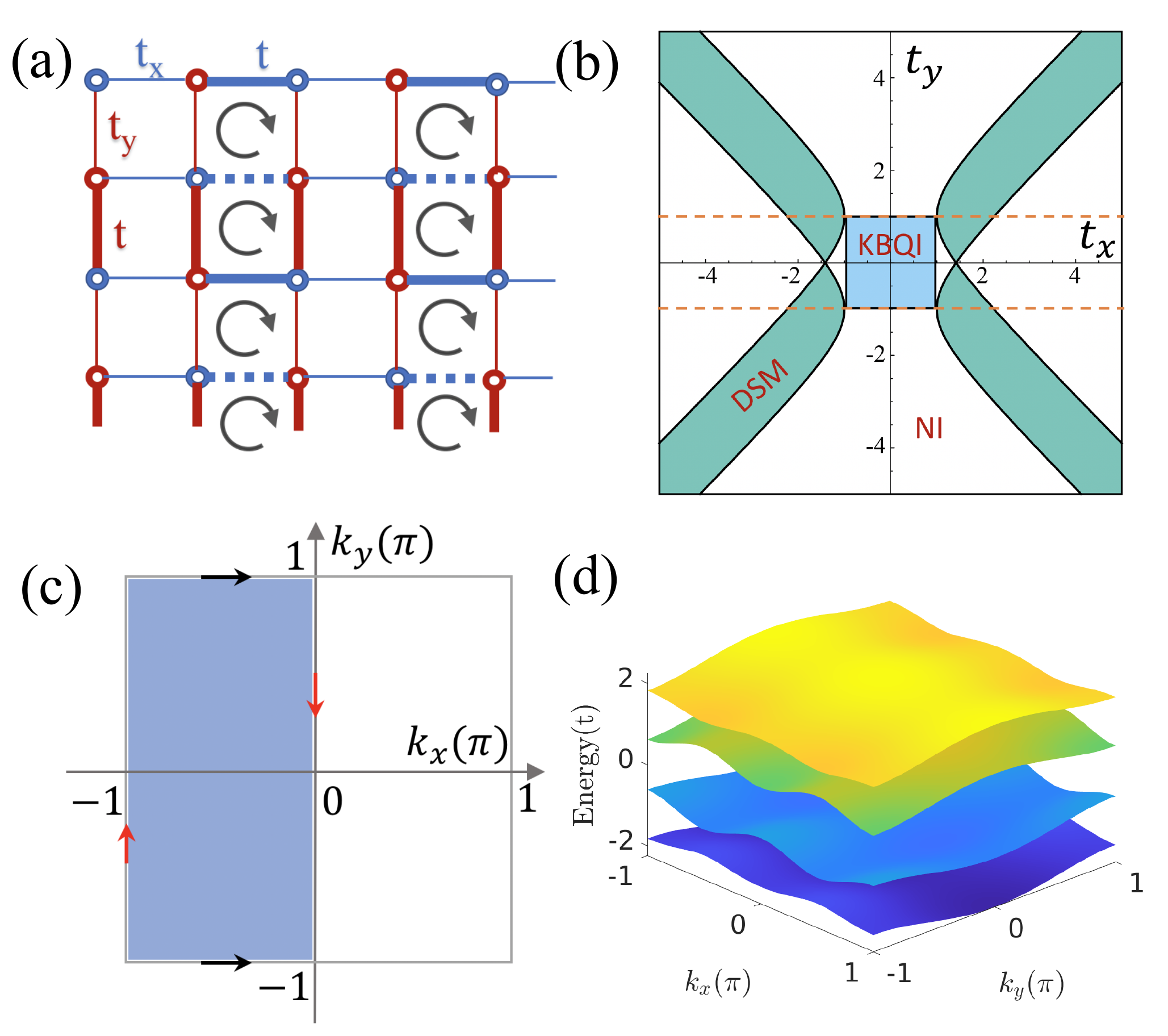}

\caption{(a) Sketch of the lattice for the first Klein-bottle BBH model with
a specific $\pi$ flux pattern. The dashed lines indicate negative
sign to account for the $\pi$ fluxes. (b) Phase diagram in the parameter
space $(t_{x},t_{y})$. The light blue region indicates the nontrivial
Klein-bottle quadrupole insulators (KBQI), the light green region
indicates the Dirac semimetal (DSM) phase. The region between two
dashed lines represents phases with nontrivial Klein-bottle topology.
Other regions are normal insulators (NI). (c) Fundamental domain of
momentum space in the BZ (blue color). The boundaries marked with
same colored arrows should be identified in that sense, thus a Klein
bottle. (d) Band structure for the first Klein-bottle BBH model in
the insulating phase. There are four non-degenerate bands. The parameters
are taken as $t_{x}=0.6$ and $t_{y}=0.3$ in units of $t$. \label{fig:model I}}
\end{figure}

We consider the first model as sketched in Fig.~\ref{fig:model I}(a).
Compared with the original BBH model, there are no uniform $\pi$
fluxes in the whole 2D lattice. The $\pi$ fluxes apply only at the
even numbered columns of plaquettes. The tight-binding Hamiltonian
reads

\begin{alignat}{1}
H_{1}= & \sum_{\mathbf{R}}\Big[t_{x}(C_{\mathbf{R},1}^{\dagger}C_{\mathbf{R},3}+C_{\mathbf{R},2}^{\dagger}C_{\mathbf{R},4})\nonumber \\
 & +t_{y}(C_{\mathbf{R},1}^{\dagger}C_{\mathbf{R},4}+C_{\mathbf{R},2}^{\dagger}C_{\mathbf{R},3})\nonumber \\
 & +(-tC_{\mathbf{R},1}^{\dagger}C_{\mathbf{R}+\hat{x},3}+tC_{\mathbf{R},4}^{\dagger}C_{\mathbf{R}+\hat{x},2})\nonumber \\
 & +t(C_{\mathbf{R},1}^{\dagger}C_{\mathbf{R}+\hat{y},4}+C_{\mathbf{R},3}^{\dagger}C_{\mathbf{R}+\hat{y},2})\Big]+\mathrm{H.c.},\label{eq:H1TB}
\end{alignat}
where $t_{x/y}$ and $t$ are the corresponding hopping amplitudes
along the $x$ and $y$ directions, as indicated in Fig.~\ref{fig:model I}(a).
The operators $C_{\mathbf{R},\zeta}^{\dagger}$ ($C_{\mathbf{R},\zeta}$)
are creation (annihilation) operators at unit cell $\mathbf{R}$ with
$\zeta\in\{1,2,3,4\}$ being orbital degrees of freedom. Note that
the minus sign from the $\pi$ fluxes is encoded in the term $-tC_{\mathbf{R},1}^{\dagger}C_{\mathbf{R}+\hat{x},3}+\mathrm{H.c.}$.
We have set the lattice constant $a=1$. In momentum space, the corresponding
Bloch Hamiltonian reads

\begin{alignat}{1}
H_{1}({\bf k})= & t_{x}\tau_{1}\sigma_{0}+[-t\cos k_{x}\gamma_{3}+t\sin k_{x}\gamma_{4}\nonumber \\
 & +(t_{y}+t\cos k_{y})\gamma_{1}-t\sin k_{y}\gamma_{2}],\label{eq:H1_k}
\end{alignat}
where $\gamma_{j}=\tau_{1}\sigma_{j}$ and $\gamma_{4}=\tau_{2}\sigma_{0}$
are the gamma matrices. The Pauli matrices $\tau$ and $\sigma$ correspond
to different orbital degrees of freedom in the unit cell, ${\bf k}=(k_{x},k_{y})$
is the momentum in 2D. Different from the original BBH model, in addition
to the four anticommuting Dirac matrices in Eq\textcolor{black}{.\ \eqref{eq:H1_k}},
there is an extra term $t_{x}\tau_{1}\sigma_{0}$. The Hamiltonian
respects chiral symmetry $\gamma_{5}H_{1}({\bf k})\gamma_{5}^{-1}=-H_{1}({\bf k})$
with the chiral symmetry operator defined as $\gamma_{5}\equiv-\gamma_{1}\gamma_{2}\gamma_{3}\gamma_{4}=\tau_{3}\sigma_{0}$.
It has time-reversal symmetry $\mathcal{T}H_{1}({\bf k})\mathcal{T}^{-1}=H_{1}(-{\bf k})$
as well, where $\mathcal{T}=K$ is just the complex conjugate operation.
Therefore, particle-hole symmetry is also preserved. 

With the help of chiral symmetry, the energy spectra can be obtained
as 
\begin{equation}
E_{\eta}^{\pm}({\bf k})=\pm\sqrt{\epsilon_{y}^{2}(k_{y})+t^{2}+t_{x}^{2}+2\eta t_{x}\sqrt{\epsilon_{y}^{2}(k_{y})+t^{2}\cos^{2}k_{x}}},\label{eq:Bandstructure1}
\end{equation}
where $\epsilon_{y}^{2}(k_{y})\equiv t_{y}^{2}+2t_{y}t\cos k_{y}+t^{2}$,
and $\eta=\pm1$. The two lower (upper) bands are no longer degenerate
unless $t_{x}=0$ {[}see Fig.~\ref{fig:model I}(d){]}. We find that
there are insulating phases as well as semimetal phases, as shown
in the phase diagram in Fig.~\ref{fig:model I}(b), different from
that of the BBH model. We focus on the insulating phases in this section
and delegate the discussion of the semimetal phases to Sec. IV. 

\subsection{Klein-bottle nontrivial phases, glide edge spectra, and Wannier bands}

Due to the gauge degrees of freedom from $\pi$ fluxes, the hopping
amplitudes are allowed to take phases $\pm1$. Thus, the $\pi$ fluxes
endow the system with a $\mathbb{Z}_{2}$ gauge field. This gauge
field can projectively modify the algebra of certain symmetry operators
\cite{YangJ18JAP}. The Klein-bottle BBH model has mirror symmetry
along $x$ direction as $\mathcal{M}_{x}H_{1}({\bf k})\mathcal{M}_{x}=H_{1}(-k_{x},k_{y})$
with $\mathcal{M}_{x}=\tau_{1}\sigma_{0}$. However, along $y$ direction,
the system does not have an exact mirror symmetry, it has a mirror
symmetry only after a gauge transformation acting on the $\mathbb{Z}_{2}$
gauge fields \cite{BBH17prb}. That is $\mathcal{M}_{y}=G(M_{y})M_{y}$
with a gauge transformation $G(M_{y})$. However, we see that the
gauge transformation $G(M_{y})$ is not compatible with the translation
operation $\mathcal{L}_{x}$. The relation between $\mathcal{M}_{y}$
and $\mathcal{L}_{x}$ becomes projectively modified as $\{\mathcal{M}_{y},\mathcal{L}_{x}\}=0$
due to $\mathbb{Z}_{2}$ gauge fields \cite{ChenZ22nc}, instead of
$[\mathcal{M}_{y},\mathcal{L}_{x}]=0$ without the gauge field. This
fundamental change of commutation relation introduces the nonsymmorphic
symmetry in momentum space and makes the fundamental domain of momentum
space a Klein bottle \cite{ChenZ22nc}. Specifically, we find 

\begin{equation}
\mathcal{M}_{y}H_{1}({\bf k})\mathcal{M}_{y}^{-1}=H_{1}(k_{x}+\pi,-k_{y}),
\end{equation}
where $\mathcal{M}_{y}=\tau_{1}\sigma_{1}$ in the chosen basis. This
corresponds to a glide-mirror symmetry in momentum space. Hence, the
momentum at $(k_{x},k_{y})$ is equivalent to $(k_{x}+\pi,-k_{y})$.
Consequently, the original BZ (torus) is reduced to two equivalent
fundamental domains {[}Klein bottles in Fig.~\ref{fig:model I}(c){]}.
In the following, we use the term \textit{Klein-bottle BZ} to indicate
the fundamental domain of momentum space. 

\begin{figure}
\includegraphics[width=1\linewidth]{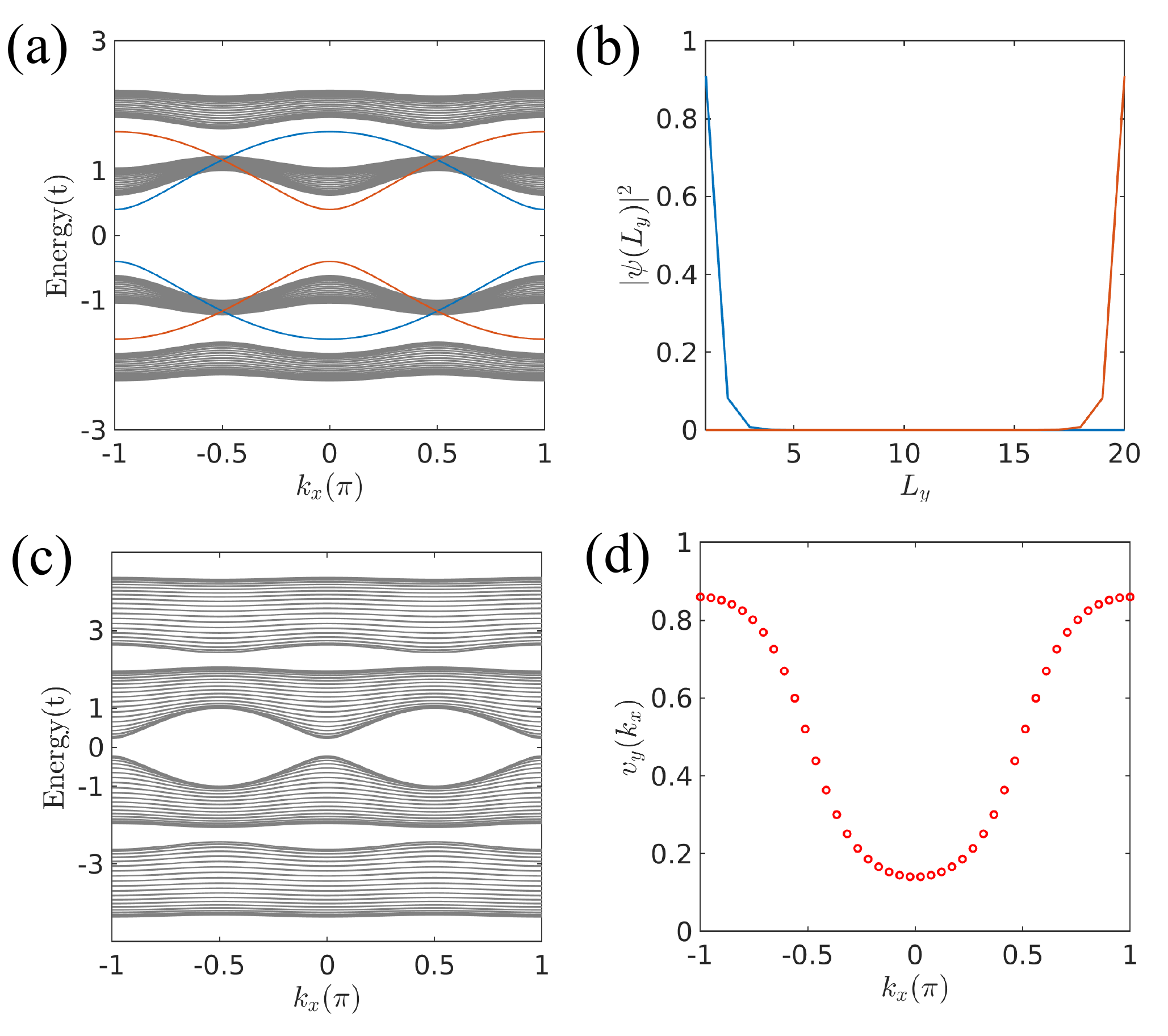}

\caption{(a) Spectra of a ribbon along $x$ direction. The blue and red color
line indicate the twined edge modes. (b) Wave function distribution
of the twined edge modes along $y$ direction of the ribbon. (c) Similar
to (a) but in a topological trivial case without edge modes. (d) The
Wannier spectrum $\nu_{y}(k_{x})$ of the lowest energy band. The
parameters are $t_{x}=0.6,t_{y}=0.3$ for (a),(b) and (d), and $t_{x}=1.2,t_{y}=2$
for (c) in units of $t$. \label{fig:Edgemodes}}
\end{figure}

Consider a ribbon geometry along $x$ direction with open boundary
conditions along $y$ direction. We find that there are edge modes
residing within the bulk bands. If we resolve their spatial distributions,
we find that the two pairs of edge modes emerge at different boundaries.
Similar to those in the BBH model, those edge modes are gapped, as
shown in the Figs.~\ref{fig:Edgemodes}(a) and (b). However, there
are essential differences. The two pairs of edge spectra have a relative
momentum shift $\delta k_{x}=\pi$, which is due to the glide-mirror
symmetry as stated above. Due to the relative momentum shift $\delta k_{x}=\pi$,
two branches of edge modes from different pairs twine around each
other from $k_{x}=-\pi$ to $k_{x}=\pi$. We call them \textit{twined
edge modes}. Moreover, the edge spectra cross the bulk continuum without
hybridization. The energy spectra of the twined edge modes can be
obtained as \cite{LiCA20prl,LuoX23prb}

\begin{alignat}{1}
E_{\mathrm{b}}(k_{x}) & =\pm\sqrt{t_{x}^{2}+t^{2}+2t_{x}t\cos(k_{x}+\theta)},
\end{alignat}
where $\theta=0/\pi$ parametrizes the two different pairs of edge
modes. 

The existence of twined edge modes can be attributed to a topological
invariant. In Ref. \cite{ChenZ22nc}, the corresponding topological
invariant is defined at the boundary of the Klein-bottle BZ as $w=\frac{1}{2\pi}[\gamma_{y}(k_{x}=0)+\gamma_{y}(k_{x}=\pi)]\ \mathrm{mod}\ 2$
where $\gamma_{y}(k_{x})$ is the Berry phase for the reduced 1D Hamiltonian
$h(k_{y})$ at a specific $k_{x}$. This topological invariant is
closely related to 1D charge polarization \cite{ChenZ22nc}. Note
that this invariant becomes ill-defined once the Klein-bottle BZ is
broken. This happens when the value of magnetic flux deviates from
$\pi$. Because in that case the relation $\{\mathcal{M}_{y},\mathcal{L}_{x}\}=0$
does not hold anymore. However, the twined edge modes, which serves
as a practical indicator of Klein-bottle insulators, may survive under
such flux deviations. 

We alternatively employ the method of Wilson loops to characterize
the topology of Klein-bottle phases. At half filling, we find that
the bulk polarization of the system vanishes. Since the two lowest
bands are not degenerate in this model and the twined edge modes reach
the gap between these two bands, we consider the polarization of the
\textit{lowest} energy band (similar results can be obtained for the
second band). We consider the ribbon geometry along $x$ direction.
Thus, the bulk polarization $p_{y}^{\kappa}$ along $y$ direction
determines the existence of twined edge modes. In the nontrivial phase
with $p_{y}^{\kappa}=\frac{1}{2}$, there are twined edge modes. In
the trivial phase with $p_{y}^{\kappa}=0$, no edge modes exist. The
polarization $p_{y}^{\kappa}$ is closely related to the Wannier center.
We obtain the Wannier center from the Wilson loop method. To this
end, we define the Wilson loop along $y$ direction at specific $k_{x}$
in the Klein-bottle BZ, i.e., $W_{y}(k_{x})$. Then the eigenvalues
of $W_{y}(k_{x})$ yield the Wannier center $\nu_{y}(k_{x})$. The
Wannier center indicates the average position of electrons relative
to the center of the unit cell. The set of Wannier centers along $y$
direction as a function of $k_{x}$ form the Wannier bands $\nu_{y}(k_{x})$.
The topological invariant for the Klein-bottle insulator can be defined
as 

\begin{align}
p_{y}^{\kappa} & =\frac{2}{L_{x}}\sum_{k_{x}=-\pi}^{0}\nu_{y}(k_{x}),
\end{align}
which is the bulk polarization of the lowest energy band. Note that
we can take the sum with respect to $k_{x}$ from $-\pi$ to $0$.
The other range from $0$ to $\pi$ can be obtained by symmetry. Here,
$L_{x}$ is the number of unit cells in $x$ direction. The topological
invariant $p_{y}^{\kappa}$ should not be changed by flux perturbations
as long as chiral symmetry is preserved. 

\begin{figure}
\includegraphics[width=0.8\linewidth]{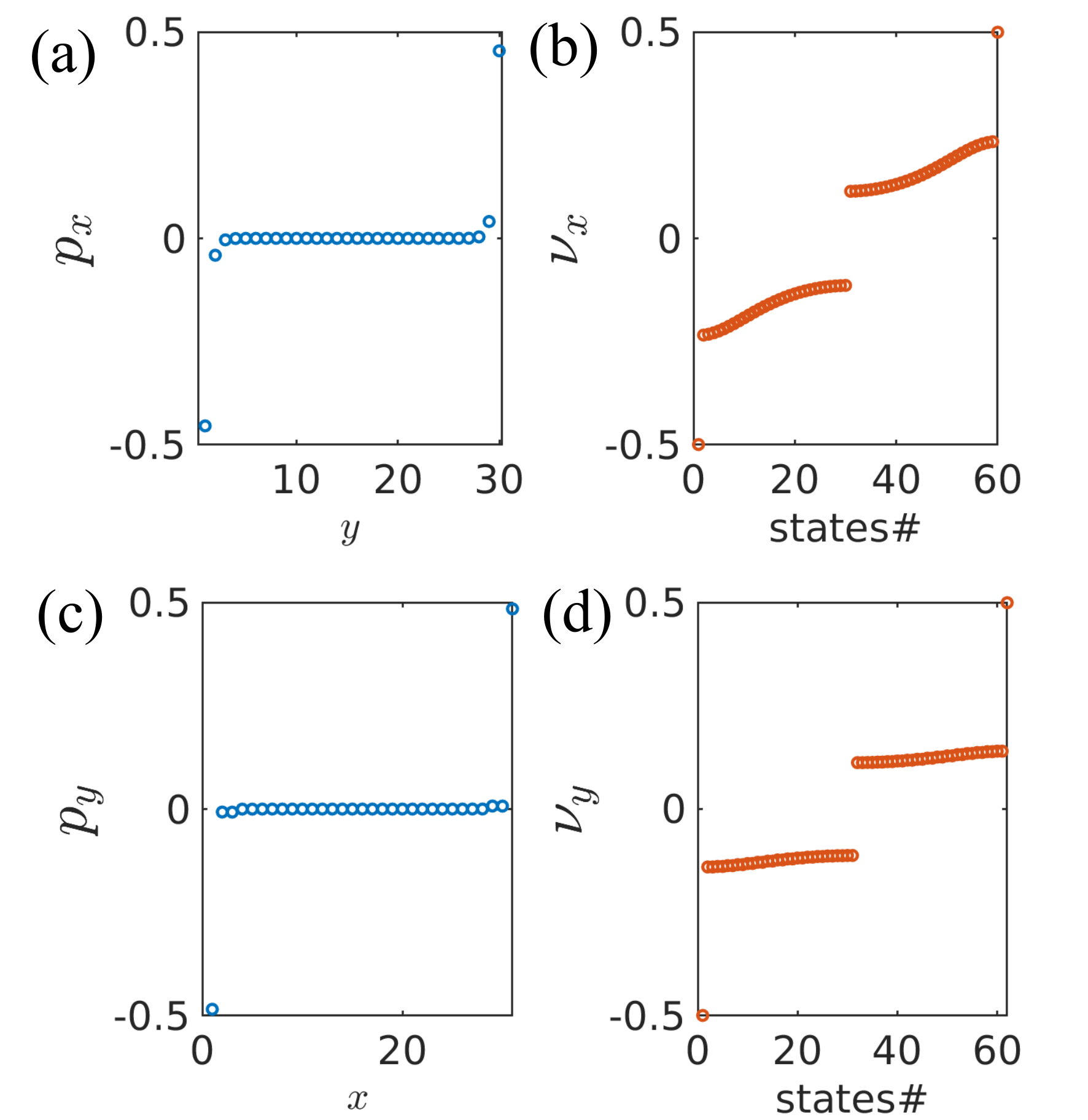}

\caption{(a) Edge polarization $p_{x}$ along $y$, (b) Wannier center $\nu_{x}$
for different eigenstates. (c) Edge polarization $p_{y}$ along $x$,
(b) Wannier center $\nu_{y}$ for different eigenstates. The parameters
are taken as $t_{x}=0.6,t_{y}=0.3$ in units of $t$. \label{fig:edgepolarization1}}
\end{figure}

The Wannier band $\nu_{y}(k_{x})$ for the lowest energy band is plotted
in Fig. \ref{fig:Edgemodes}(d). For the nontrivial Klein-bottle insulator
phase, the Wannier band $\nu_{y}(k_{x})$ has to cross $\nu_{y}=\frac{1}{2}$.
Due to the periodicity of the BZ, $\nu_{y}(k_{x})$ crosses $\nu_{y}=\frac{1}{2}$
an even number of times. Therefore, we obtain the bulk polarization
$p_{y}^{\kappa}=\frac{1}{2}$ from the lowest energy bands as a topological
invariant for Klein-bottle insulators. Note that the Wannier band
crosses $\nu_{y}=\frac{1}{2}$ once within the domain $k_{x}\in[0,\pi]$,
consistent with the winding number defined in Ref. \cite{ChenZ22nc}.
For the trivial insulator case, it does not cross the value $\nu_{y}=\frac{1}{2}$
at all, thus $p_{y}^{\kappa}=0$. 

We emphasize that at $k_{x}=\pm\frac{\pi}{2}$, the Wannier center
is fixed at $0$ or $\frac{1}{2}$. This is because at these special
points the Hamiltonian $H_{1}(\pm\frac{\pi}{2},k_{y})$ has space-time
inversion symmetry, which can quantize the Wannier center \cite{ChiuCK18arxiv}.
From the topological invariant $p_{y}^{\kappa}$, we find that the
nontrivial Klein-bottle phase exists for 
\begin{equation}
|t_{y}|<1\ \ \Rightarrow\text{\ensuremath{p_{y}^{\kappa}}\ensuremath{=\ensuremath{\frac{1}{2}}}, }
\end{equation}
as indicated in Fig.~\ref{fig:model I}(b). Note that this phase
regime contains Klein-bottle insulators (insulating phase characterized
by twined edge modes) and as well as Klein-bottle Dirac semimetals
(semimetals with twined edge modes). 

\begin{figure}
\includegraphics[width=1\linewidth]{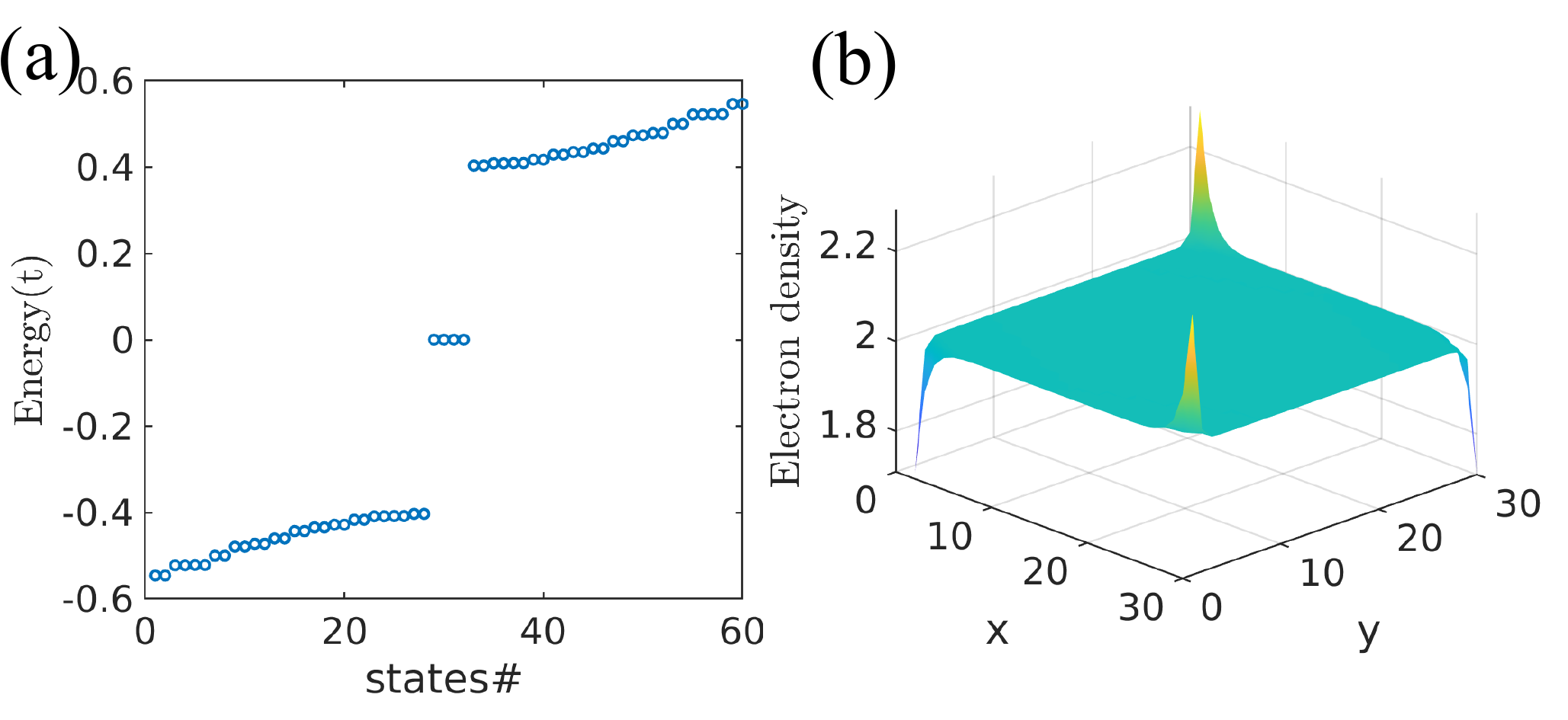}

\caption{(a) Energy spectrum of the model with open boundaries in both directions.
There are four zero-energy corner modes in the spectrum. (b) Electron
charge density distribution on the lattice. It gives fractional corner
charges $\pm\frac{e}{2}$. The parameters are taken as $t_{x}=0.6,t_{y}=0.3$
in units of $t$. \label{fig:cornermodes1}}
\end{figure}

\subsection{Nontrivial Klein-bottle quadrupole insulators}

Twined edge modes exist in the nontrivial Klein-bottle phases as a
consequence of first-order topology. In our model, the twined edge
modes are gapped as well. However, there is a relative momentum shift
of the spectra at different edges. These spectra can touch, cross
or be hidden in the bulk continuum. An intriguing question is whether
such twined edge modes can support second-order topology, characterized
by corner states and fractional charges. 

To characterize the system, we first calculate the quadrupole moment
$q_{xy}$. Afterwards, we check the corresponding edge and corner
signatures. The quadrupole moment can be obtained in real space as
\cite{Wheeler19prb,Kang19prb,Roy19prr,LiCA20prl,YangYB21prb},
\begin{equation}
q_{xy}=\frac{1}{2\pi}\mathrm{Imlog}\left[\mathrm{det}(U^{\dagger}\hat{Q}U)\sqrt{\mathrm{det}(Q^{\dagger})}\right],
\end{equation}
where $\hat{Q}\equiv\exp[i2\pi\hat{q}_{xy}]$ with $\hat{q}_{xy}=\hat{x}\hat{y}/(L_{x}L_{y})$
being quadrupole momentum density operator per unit cell at position
${\bf R}=(x,y)$. Here, $\hat{x}(\hat{y})$ is the position operator
along $x$ ($y$) direction and $L_{x(y)}$ is the corresponding system
size. The matrix $U$ is constructed by packing all the occupied eigenstates
in a column-wise way. The quantization of $q_{xy}$ is protected by
chiral symmetry \cite{LiCA20prl,YangYB21prb}. For an insulating phase,
it is a nontrivial quadrupole insulator when $q_{xy}=\frac{1}{2}$.
We find the nontrivial Klein-bottle quadrupole insulator phase in
the regime $|t_{x}|<1$ and $|t_{y}|<1$ {[}see Fig.~\ref{fig:model I}(b){]},
similar to the BBH model. 

The edge polarizations $p_{x}^{\mathrm{edge}}$ and $p_{y}^{\mathrm{edge}}$
can also help to detect the topologically nontrivial phase. Take $p_{x}^{\mathrm{edge}}$
as an example. Consider a ribbon along $x$ direction with width $L_{y}$
along $y$ direction. Employing the Wilson loop method, the edge polarization
is calculated as \cite{Benalcazar17Science,BBH17prb,LiC20prb}

\begin{equation}
p_{x}^{\mathrm{edge}}=\sum_{y=1}^{L_{y}/2}p_{x}(y),
\end{equation}
where $p_{x}(y)$ is the distribution of polarization along $y$ direction.
We calculate this spatial-resolved polarization as 
\begin{equation}
p_{x}(y)=\sum_{j=1}^{2L_{y}}\rho^{j}(y)\nu_{y}^{j}(k_{x}),
\end{equation}
where $\rho^{j}(y)=\frac{1}{L_{x}}\sum_{k_{x},\zeta}|\sum_{n}[u_{k_{x}}^{n}]^{y,\zeta}[\nu_{k_{x}}^{j}]^{n}|^{2}$.
Here, $[\nu_{k_{x}}^{j}]^{n}$ is the $n$-th component of the $j$-th
Wilson-loop eigenstate $|\nu_{k_{x}}^{j}\rangle$ corresponding to
the eigenvalue $\nu_{y}^{j}(k_{x})$, while $[u_{k_{x}}^{n}]_{y,\zeta}$
is the $n$-th eigenstate of the Hamiltonian $H_{y}(k_{x})$ on the
ribbon with integer number $n\in\{1,2,3,...,2L_{y}\}$. The edge polarization
$p_{y}^{\mathrm{edge}}$ in $y$ direction can be calculated in a
similar way. For a nontrivial quadrupole insulator, $(p_{x}^{\mathrm{edge}},p_{y}^{\mathrm{edge}})=(\frac{1}{2},\frac{1}{2})$. 

Consider a ribbon along $x$ direction with open boundaries in $y$
direction. In Figs. \ref{fig:edgepolarization1}(a) and \ref{fig:edgepolarization1}(b),
two topological states are localized on the edge with half-integer
Wannier values, while the other states are distributed over the bulk.
The edge polarization $p_{x}(y)$ becomes nonzero at the sample edge.
The spatial-resolved polarization yields quantized edge polarization
$p_{x}^{\mathrm{edge}}=\frac{1}{2}$ ($p_{y}^{\mathrm{edge}}=\frac{1}{2}$).
Similar results appear for a ribbon along $y$ direction, as shown
in Figs. \ref{fig:edgepolarization1}(c) and \ref{fig:edgepolarization1}(d). 

Moreover, there are zero-energy corner modes carrying fractional corner
charges. We show in Fig. \ref{fig:cornermodes1}(a) that there are
four-fold degenerate zero-energy modes, whose wave functions are sharply
localized at corners of the sample. It is also found that the corner
charges are fractionalized at $\pm e/2$ {[}see Fig. \ref{fig:cornermodes1}(b){]}.

\begin{figure}
\includegraphics[width=1\linewidth]{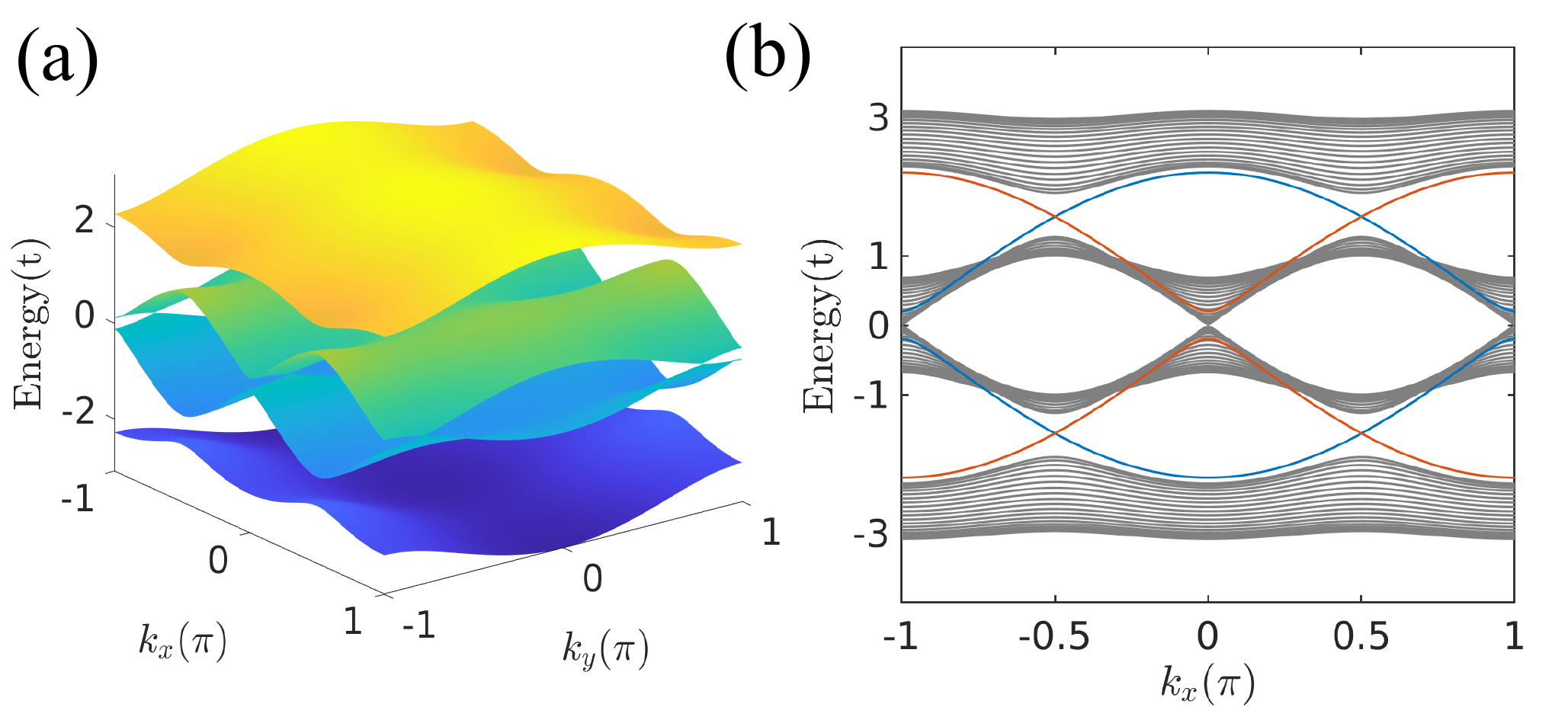}

\caption{(a) Four Dirac points in the band structure. (b) Coexistence of edge
Dirac points and bulk edge Dirac points in the spectra of a ribbon
along $x$ direction. The parameters are taken as $t_{x}=1.2,t_{y}=0.6$
in units of $t$. \label{fig:DiracSM1}}
\end{figure}

\section{Klein-bottle Dirac semimetals on square lattices}

Beside the Klein-bottle insulating phase, a Dirac semimetal phase
also exists, as shown in Fig. \ref{fig:DiracSM1}. There are four
Dirac points residing in the BZ. From the energy band solutions in
Eq\textcolor{black}{.\ \eqref{eq:Bandstructure1}}, to obtain band
touching at zero energy (due to chiral symmetry), we require $\cos^{2}k_{x}=1$,
i.e., $k_{x}\in\{0,\pi\}$. In this case, the energy spectra are simplified
as 

\begin{equation}
E_{\eta}^{\pm}({\bf k})=\pm\left(\sqrt{\epsilon_{y}^{2}(k_{y})+t^{2}}+2\eta t_{x}\right),\label{eq:Bandstructure1-S}
\end{equation}
Therefore, four Dirac points are located at 

\begin{equation}
(K_{x},K_{y})=\left(0/\pi,\ \ \pm\arccos\left[\frac{t_{x}^{2}-t_{y}^{2}-2t^{2}}{2t_{y}t}\right]\right).
\end{equation}
The valid solutions for $K_{y}$ give rise to the Klein-bottle Dirac
semimetal phase, determined by the overlap of two hyperbolas in the
parameter space $(t_{x},t_{y})$ as 

\begin{equation}
(t_{y}\pm t)^{2}-t_{x}^{2}=t^{2},
\end{equation}
as shown in the Fig. \ref{fig:model I}(b). The Dirac semimetal phase
that appears in this model has no counterpart in the original BBH
model. 

The Dirac points are located at the boundary of the Klein-bottle BZ.
They come in two dual pairs related by glide-mirror symmetry $\mathcal{M}_{y}$
as $\{(0,\pm K_{y})\Leftrightarrow(\pi,\mp K_{y})\}$. We know that
the topological protection of Dirac points is typically related to
a winding number defined on a path enclosing the Dirac points. Due
to chiral symmetry, rewriting the Hamiltonian Eq\textcolor{black}{.\ \eqref{eq:H1_k}}
in an off-diagonal form, this leads to

\begin{equation}
H_{1}({\bf k})=\left(\begin{array}{cc}
0 & q({\bf k})\\
q^{\dagger}({\bf k}) & 0
\end{array}\right),
\end{equation}
where 
\begin{alignat}{1}
q({\bf k}) & =\left(\begin{array}{cc}
-t_{x}+te^{ik_{x}} & t_{y}+te^{ik_{y}}\\
t_{y}+te^{-ik_{y}} & t_{x}+te^{-ik_{x}}
\end{array}\right).
\end{alignat}
The winding number for the Dirac points is defined as $\omega=\frac{1}{2\pi i}\oint_{\ell}d{\bf k}\cdot\mathrm{Tr}[q^{-1}({\bf k})\nabla_{{\bf k}}q({\bf k})]$
\cite{Schnyder08prb,ChiuCK16rmp}, where the loop $\ell$ is chosen
such that it encloses a single Dirac point. 

The twined edge modes from nontrivial Klein-bottle topology also appear
in the Dirac semimetal phase, as shown in the Fig. \ref{fig:DiracSM1}(b).
The bulk Dirac points coexist with edge Dirac points, located at different
energies. For certain parameters, they are hidden in the bulk bands
but not directly merge with the continuum. From the wave function
distribution, we find that the edge modes are well-localized at boundaries
even if they coexist with bulk bands. In the Dirac semimetal phase
with a trivial Klein-bottle topology for $|t_{y}|>1$, the twined
edge modes disappear. 

\begin{figure}
\includegraphics[width=1\linewidth]{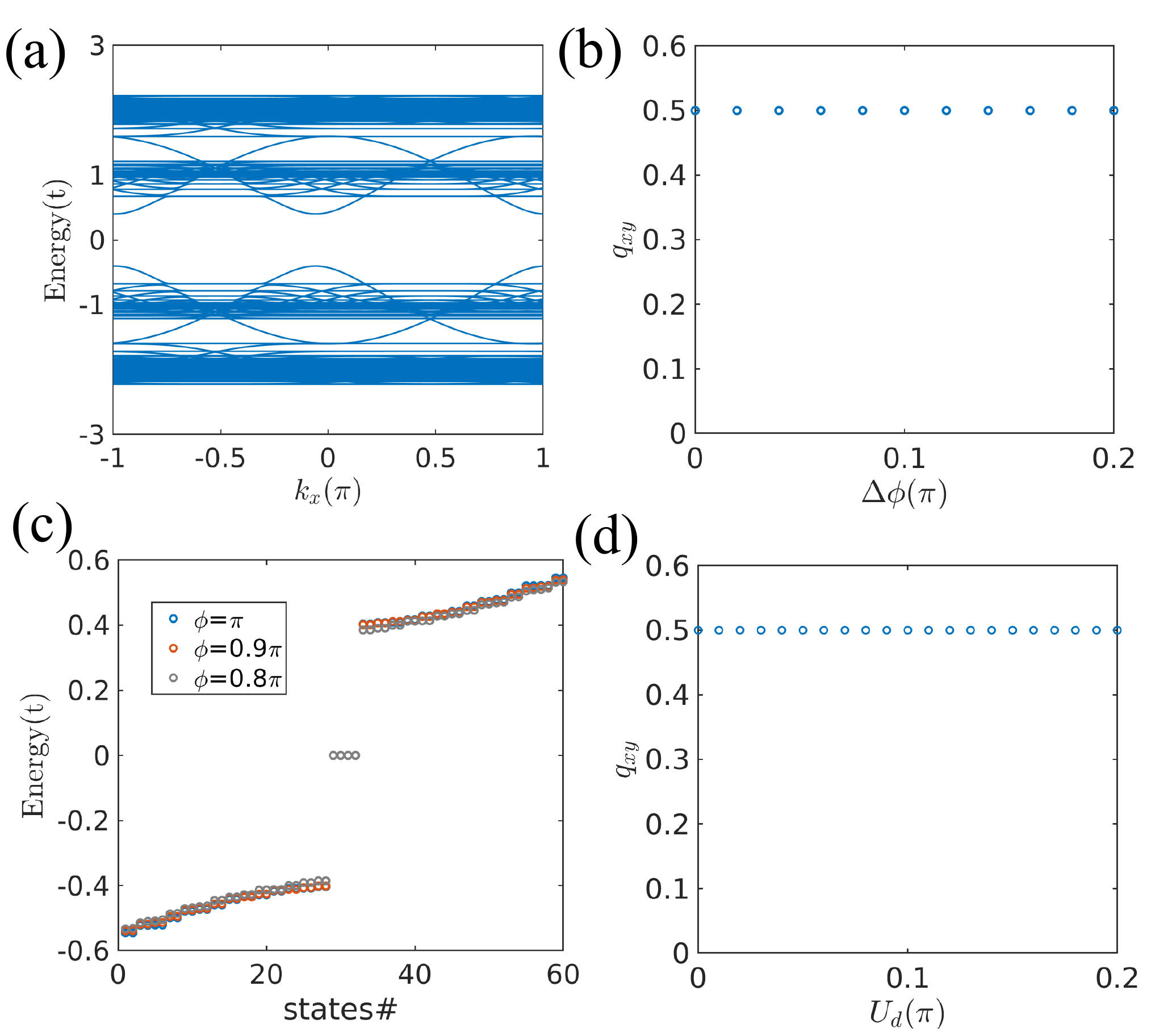}

\caption{(a) Energy spectrum on a ribbon along $x$ direction when $\phi=0.95\pi$.
(b) Quadrupole moment $q_{xy}$ as a function of flux deviation $\Delta\phi=\pi-\phi$.
(c) Eigenstates around zero energy for different $\Delta\phi$. (d)
Quadrupole moment $q_{xy}$ as a function of random flux strength
$U_{d}$. The parameters are taken as $t_{x}=0.6$, $t_{y}=0.3$ in
units of $t$. \label{fig:fluxrobust}}
\end{figure}

\section{Robustness of nontrivial quadrupole insulators against flux perturbations}

In the previous section, we showed that the realization of Klein-bottle
quadrupole insulators relies on exact threading of $\pi$ fluxes on
even numbers of plaquettes in the 2D lattice. We now address the question
of how robust nontrivial Klein-bottle quadrupole insulators are against
flux deviations from the value of $\pi$. We check the stability of
the nontrivial Klein-bottle quadrupole insulators in two cases: In
the first one, the flux $\phi$ deviates from $\pi$ but is still
uniform in the lattice. This helps us to determine how general the
nontrivial phases are. In the second one, the flux value is chosen
randomly, fluctuating around $\pi$. 

\subsection{Uniform flux deviations}

Let us first check the evolution of twined edge modes for a model
with flux $\phi$ deviating from $\pi$ uniformly. In the original
case, the well-localized twined edge modes cross the bulk bands without
hybridization {[}Fig. \ref{fig:Edgemodes}(a){]}. When $\phi=0.95\pi$,
the twined edge modes almost keep their form {[}Fig. \ref{fig:fluxrobust}(a){]}
but the overlapping parts start to hybridize. In this case, the Klein-bottle
BZ manifold is broken because the glide-mirror symmetry does not hold
anymore. We define $\Delta\phi=\pi-\phi$. As $\Delta\phi$ increases
further, the twined edge modes hybridize with the other bands stronger.
This may be explained as the hybridization of Landau levels (flat
bands) and twined edge modes. 

Now, we check the robustness of Klein-bottle quadrupole insulators
against flux perturbations. To this end, we employ the quadrupole
moments $q_{xy}$ and the corresponding corner states. To be specific,
we take $t_{x}=0.6$ and $t_{y}=0.3$ in the calculations. It is demonstrated
in Fig. \ref{fig:fluxrobust}(b) that $q_{xy}$ stays at $q_{xy}=\frac{1}{2}$
even if $\Delta\phi$ grows to a relatively large value. This result
indicates the Klein-bottle quadrupole insulator is robust against
flux deviations. It also suggests that nontrivial quadrupole insulators
exist in an extended range of flux values $\phi$, not just at $\phi=\pi$.
Correspondingly, there are four zero-energy corner states in the gap
of the spectrum {[}see Fig. \ref{fig:fluxrobust}(c){]} with quantized
fractional charge at each corner. The robustness of nontrivial Klein-bottle
quadrupole insulators can be attributed to the persistence of twined
edge modes under flux perturbations. They remain almost intact and
gapped {[}Fig. \ref{fig:fluxrobust}(a){]}. 

\subsection{Random flux}

The magnetic flux can also be chosen randomly at each plaquette \cite{LiCA22prb}.
We assume that the magnetic flux $\phi$ fluctuates around $\pi$.
The flux deviation at each plaquette takes a random value from the
uniformly distributed range $[-U_{d},U_{d}]$, with $U_{d}$ being
a disorder strength. We also check the quadrupole moments $q_{xy}$
under random flux. From Fig. \ref{fig:fluxrobust}(d), $q_{xy}$ remains
at $\frac{1}{2}$ as $U_{d}$ increases from $0$ up to $0.2\pi$.
When calculating $q_{xy}$, we can also investigate a single disorder
configuration. We then find that the zero-energy midgap states and
corner charges remain robust. Together with $q_{xy}$, these findings
suggest the strong robustness of nontrivial quadrupole insulators
against random flux. 

\section{Trivial Klein-bottle quadrupole insulators with corner states}

\begin{figure}
\includegraphics[width=1\linewidth]{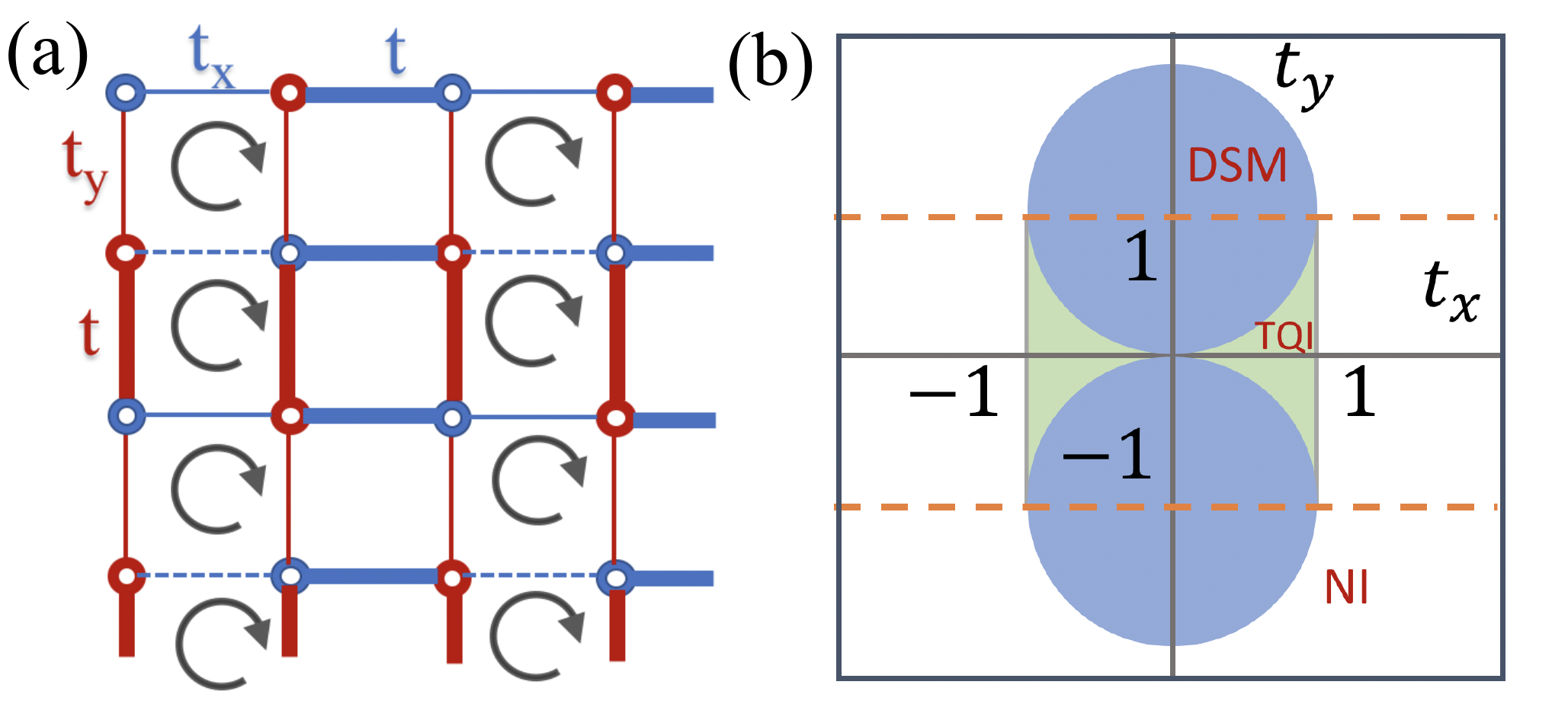}

\caption{(a) Sketch of the lattice for the second Klein-bottle BBH model with
a different $\pi$-flux pattern. The dashed lines indicate negative
sign to account for the $\pi$ fluxes. (b) Phase diagram in the parameter
space $(t_{x},t_{y})$. The light blue region indicates the Dirac
semimetal (DSM) phase, the light green region indicate the trivial
quadrupole insulators (TQI). The region between two dashed lines represent
phases with nontrivial Klein-bottle topology. Other regions are the
normal insulators (NI). \label{fig:model2}}
\end{figure}

Now, we consider the second Klein-bottle BBH model as sketched in
Fig. \ref{fig:model2}(a). The $\pi$ fluxes apply only to the odd
number of columns of plaquettes, instead of the even number of columns.
This subtle change of applying $\pi$-flux patterns makes a strong
difference. It gives rise to a totally different Hamiltonian. The
nontrivial quadrupole insulators do not appear anymore. There are
also insulators and Dirac semimetals in the phase diagram {[}Fig.
\ref{fig:model2}(b){]}, but the insulator is a trivial insulator
with $q_{xy}=0$, although it supports corner charges. 

\begin{figure}
\includegraphics[width=0.8\linewidth]{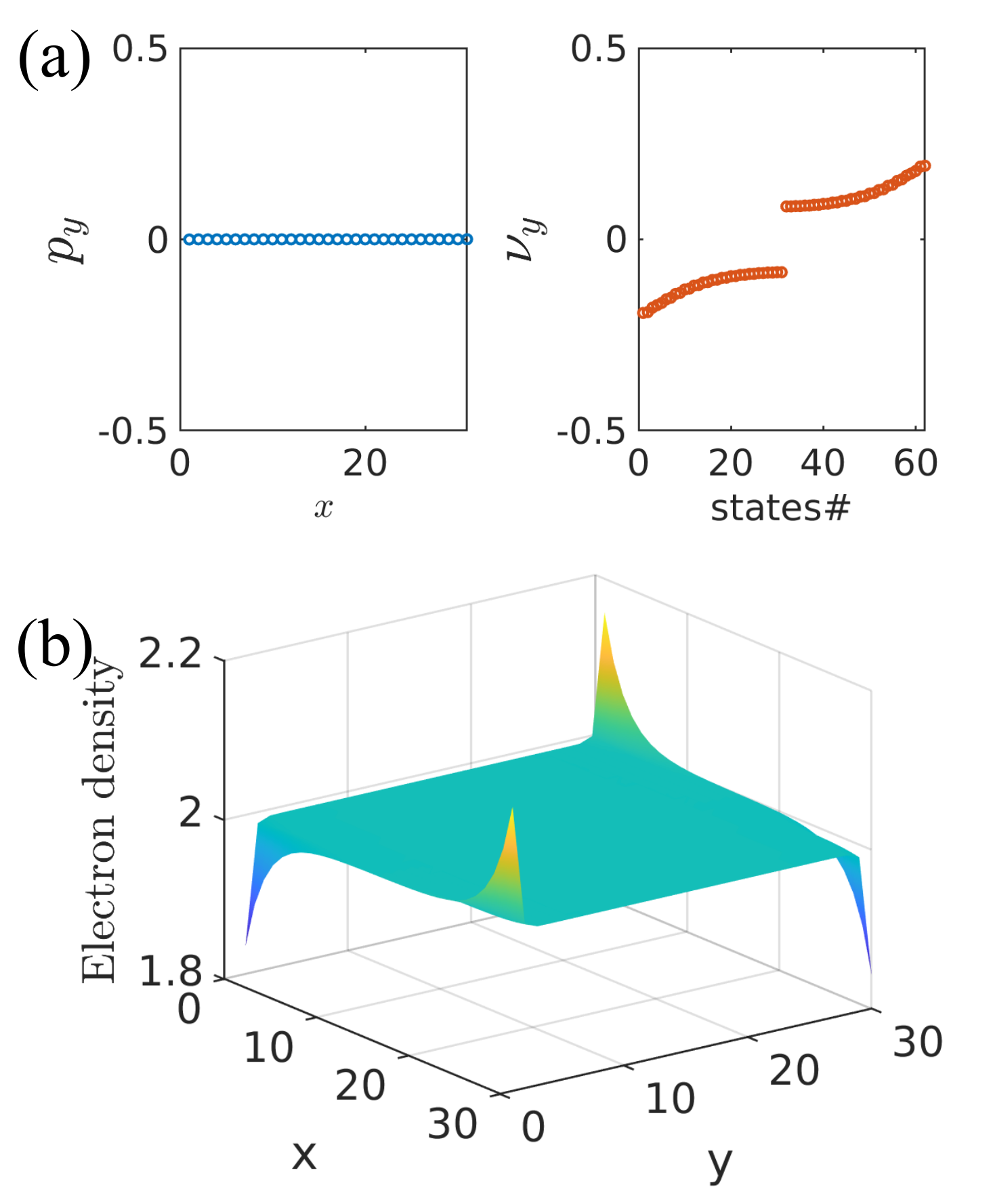}

\caption{(a) Edge polarization $p_{y}$ along $x$ (left panel) and Wannier
center $\nu_{y}$ for different eigenstates (right panel) in the second
Klein-bottle BBH model. (b) Electron density distribution in the lattice.
The parameters are taken $t_{x}=0.8,t_{y}=0.2$ in units of $t$ for
all plots. \label{fig:edgepolarize2}}
\end{figure}

The tight-binding Hamiltonian of the second Klein-bottle BBH model
reads

\begin{alignat}{1}
H_{2}= & \sum_{\mathbf{R}}\Big[(-t_{x}C_{\mathbf{R},1}^{\dagger}C_{\mathbf{R},3}+t_{x}C_{\mathbf{R},2}^{\dagger}C_{\mathbf{R},4})\nonumber \\
 & +t_{y}(C_{\mathbf{R},1}^{\dagger}C_{\mathbf{R},4}+C_{\mathbf{R},2}^{\dagger}C_{\mathbf{R},3})\nonumber \\
 & +t(C_{\mathbf{R},1}^{\dagger}C_{\mathbf{R}+\hat{x},3}+C_{\mathbf{R},4}^{\dagger}C_{\mathbf{R}+\hat{x},2})\nonumber \\
 & +t(C_{\mathbf{R},1}^{\dagger}C_{\mathbf{R}+\hat{y},4}+C_{\mathbf{R},3}^{\dagger}C_{\mathbf{R}+\hat{y},2})\Big]+\mathrm{H.c.},\label{eq:H2TB}
\end{alignat}
where the minus sign from the $\pi$ fluxes is taken into account
in the term $-t_{x}C_{\mathbf{R},1}^{\dagger}C_{\mathbf{R},3}+\mathrm{H.c.}.$
In momentum space, the Bloch Hamiltonian corresponding to Eq\textcolor{black}{.\ \eqref{eq:H2TB}}
becomes

\begin{alignat}{1}
H_{2}({\bf k})= & -t_{x}\tau_{1}\sigma_{3}+t\cos k_{x}\tau_{1}\sigma_{0}-t\sin k_{x}\tau_{2}\sigma_{3}\nonumber \\
 & +(t_{y}+t\cos k_{y})\tau_{1}\sigma_{1}-t\sin k_{y}\tau_{1}\sigma_{2}.\label{eq:H2k}
\end{alignat}
The bulk spectrum of Eq\textcolor{black}{.\ \eqref{eq:H2k}} reads

\begin{equation}
E_{\eta}^{\pm}({\bf k})=\pm\sqrt{\epsilon_{y}^{2}(k_{y})+t^{2}+t_{x}^{2}+2\eta t\sqrt{\epsilon_{y}^{2}(k_{y})+t_{x}^{2}\cos^{2}k_{x}}},\label{eq:Bandstructure1-1}
\end{equation}
where $\epsilon_{y}^{2}(k_{y})\equiv t_{y}^{2}+2t_{y}t\cos k_{y}+t^{2}$
is defined in the same way as before. This model also respects chiral
symmetry $\gamma_{5}H_{1}({\bf k})\gamma_{5}^{-1}=-H_{1}({\bf k})$.
The $\pi$-flux gauge field gives rise to nonsymmorphic symmetry in
momentum space as

\begin{equation}
\mathcal{M}_{y}^{'}H_{2}({\bf k})\mathcal{M}_{y}^{'-1}=H_{2}(k_{x}+\pi,-k_{y}),
\end{equation}
where $\mathcal{M}_{y}^{'}=\tau_{2}\sigma_{2}$ in the chosen basis. 

The phase diagram of the second Klein-bottle BBH model is plotted
in Fig. \ref{fig:model2}(b). The Dirac semimetal phase is located
inside the two circles in parameter space $(t_{x},t_{y})$

\begin{equation}
t_{x}^{2}+(t_{y}\pm t)^{2}=t^{2}.
\end{equation}
The Dirac points are at 
\begin{equation}
(K_{x},K_{y})=\left(0/\pi,\ \ \pm\arccos\left[-\frac{t_{x}^{2}+t_{y}^{2}}{2t_{y}t}\right]\right).
\end{equation}
In the Klein-bottle Dirac semimetal phases, there are twined edge
modes on a ribbon geometry with open boundary. The other regions are
insulating phases. The nontrivial Klein-bottle phase is bounded by
$|t_{y}|<1$. Compared with the first model in Eq.\ \eqref{eq:H1_k},
the different $\pi$-flux pattern in the second Klein-bottle BBH model
leads to totally different matrix structures in Eq.\ \eqref{eq:H2k}.
Thus, the corresponding energy bands are quite different, giving rise
to significantly different phase diagrams. Moreover, we notice the
exchange of variables $t\longleftrightarrow t_{x}$ in the energy
bands compared to Eq.\ \eqref{eq:Bandstructure1}. This can be effectively
viewed as the exchange of dimerized hopping strength of the SSH model
along $x$ direction, which makes the topological properties different
from those of the first model.\textcolor{blue}{{} }

In an insulator phase, we find the edge polarizations take the values
$(p_{x}^{\mathrm{edge}},p_{y}^{\mathrm{edge}})=(\frac{1}{2},0)$.
This anisotropic property of edge polarizations bears similarity to
weak topological insulators. In Fig. \ref{fig:edgepolarize2}(a),
we plot the edge polarizations $p_{y}^{\mathrm{edge}}$, together
with the Wannier values of eigenstates. The edge polarization $p_{y}^{\mathrm{edge}}$
is zero {[}the nontrivial $p_{x}^{\mathrm{edge}}=\frac{1}{2}$ is
not shown here{]}. 

Consider open boundary conditions along both $x$ and $y$ directions.
Then, the edge polarizations are terminated at corners. This leads
to charges $Q^{\mathrm{corner}}=\pm\frac{e}{2}$ localized at the
corners {[}see Fig. \ref{fig:edgepolarize2}(b){]}. There are four
zero-energy midgap states in the energy spectra when $(t_{x},t_{y})$
is located in the light green region of the phase diagram shown in
Fig. \ref{fig:model2}(b). However, if we calculate the topological
invariant $q_{xy}$, we find that the second Klein-bottle BBH model
is a trivial Klein-bottle insulator. Remarkably, it exhibits twined
edge modes (first-order) and corner-localized charges, but it has
a vanishing quadrupole moment $q_{xy}=0$ (second-order topological
invariant). The defining properties of a quadrupole insulator $|q_{xy}|=|p_{x}^{\mathrm{edge}}|=|p_{y}^{\mathrm{edge}}|=|Q^{\mathrm{corner}}|$
are not satisfied \cite{BBH17prb}. The corner charges follow $Q^{\mathrm{corner}}=p_{x}^{\mathrm{edge}}+p_{y}^{\mathrm{edge}}$.
Thus the corner charges and edge polarizations are pure surface effects,
unrelated to bulk quadrupole moments \cite{BBH17prb}. 

\begin{figure}
\includegraphics[width=1\linewidth]{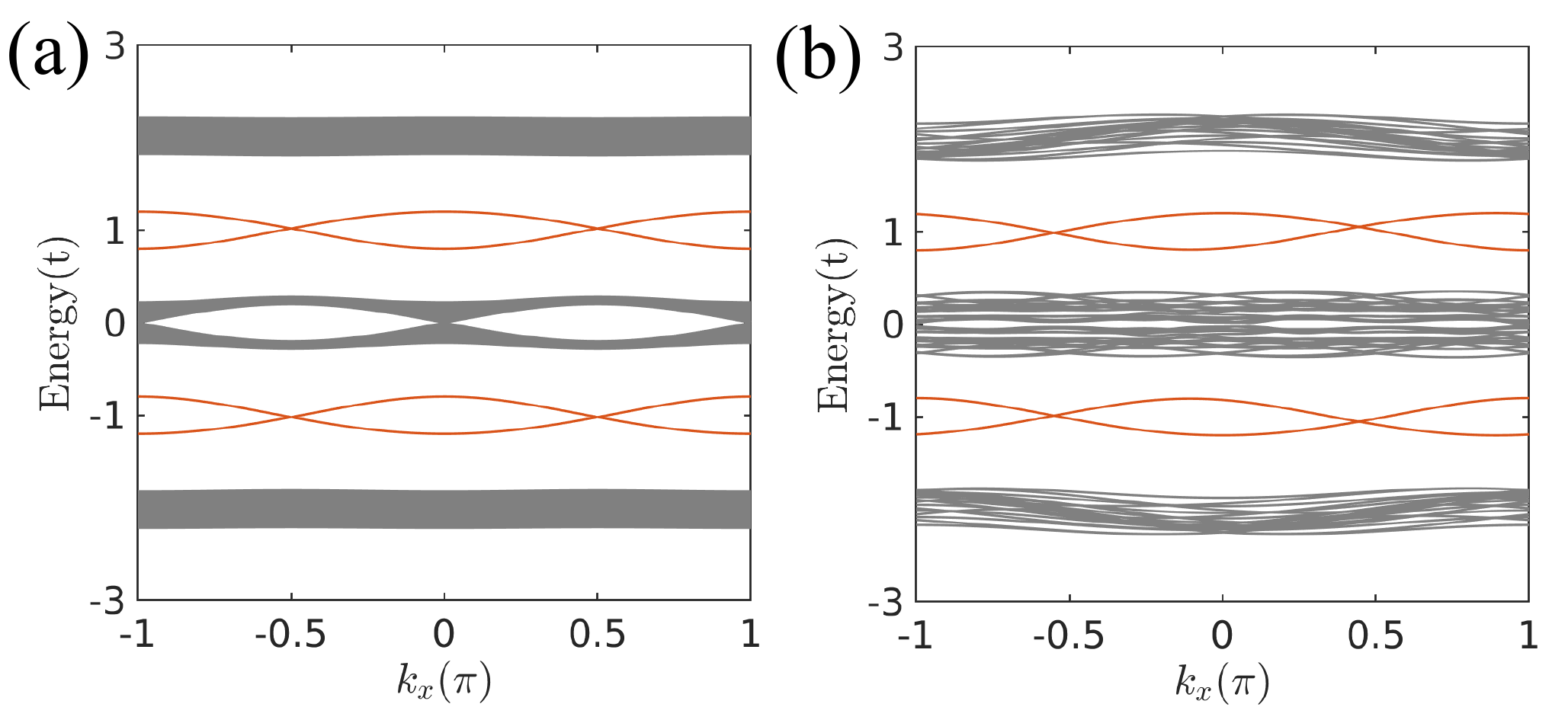}

\caption{(a) Energy spectrum of the second Klein-bottle BBH model on a ribbon
along $x$ direction. The edge and bulk Dirac points coexist. (b)
Same as panel (a) but with the flux $\phi=0.9\pi$. The parameters
are taken $t_{x}=t_{y}=0.2$ in units of $t$. \label{fig: robustness2}}
\end{figure}

We further analyze the robustness of twined edge modes in the Klein-bottle
phases when the magnetic flux deviates from $\pi$. Consider a ribbon
along $x$ direction. In the Klein-bottle Dirac semimetal phases,
the twined edge modes reside between the bulk bands and can be detached
from them. When we gradually change $\phi$ from $\pi$, the twined
edge modes persist and are detached from the bulk modes even up to
a relatively large $\Delta\phi$, as shown in Fig. \ref{fig: robustness2}(b). 

\section{discussion and conclusions}

We show that the variation of $\pi$-flux patterns changes the topology
of the considered system dramatically. Hence, particular $\pi$-flux
patterns may help to search for novel topological phases. The Klein-bottle
quadrupole insulator requires only half of the total $\pi$ fluxes
as compared to the original BBH model, simplifying experimental realizations
of nontrivial quadrupole insulators. The manipulation of magnetic
flux is possible in different synthetic systems. Therefore, our predictions
are experimentally relevant. 

In summary, we have proposed the existence of nontrivial Klein-bottle
quadrupole insulators and Dirac semimetal in 2D. The twined edge modes,
which support the second-order topology, appear as a characteristic
signature of Klein-bottle systems. We have verified the robustness
of the nontrivial quadrupole insulators against flux perturbations.
In the Klein-bottle Dirac semimetal phases, we discover the the coexistence
of edge Dirac points with bulk Dirac points. 

\section{Acknowledgments}

C.A.L. thanks Jan Budich and Bo Fu for helpful discussion. This work
has been financially supported by the Würzburg-Dresden Cluster of
Excellence ct.qmat, Project-id 390858490, the DFG (SFB 1170), and
the Bavarian Ministry of Economic Affairs, Regional Development and
Energy within the High Tech Agenda Project ``Bausteine für das Quanten
Computing auf Basis Topologischer Materialen''. H. G. acknowledges
support from the NSFC grant No. 12074022.

\appendix

\section{Revisit of the Klein-bottle insulator model}

\begin{figure}
\includegraphics[width=1\linewidth]{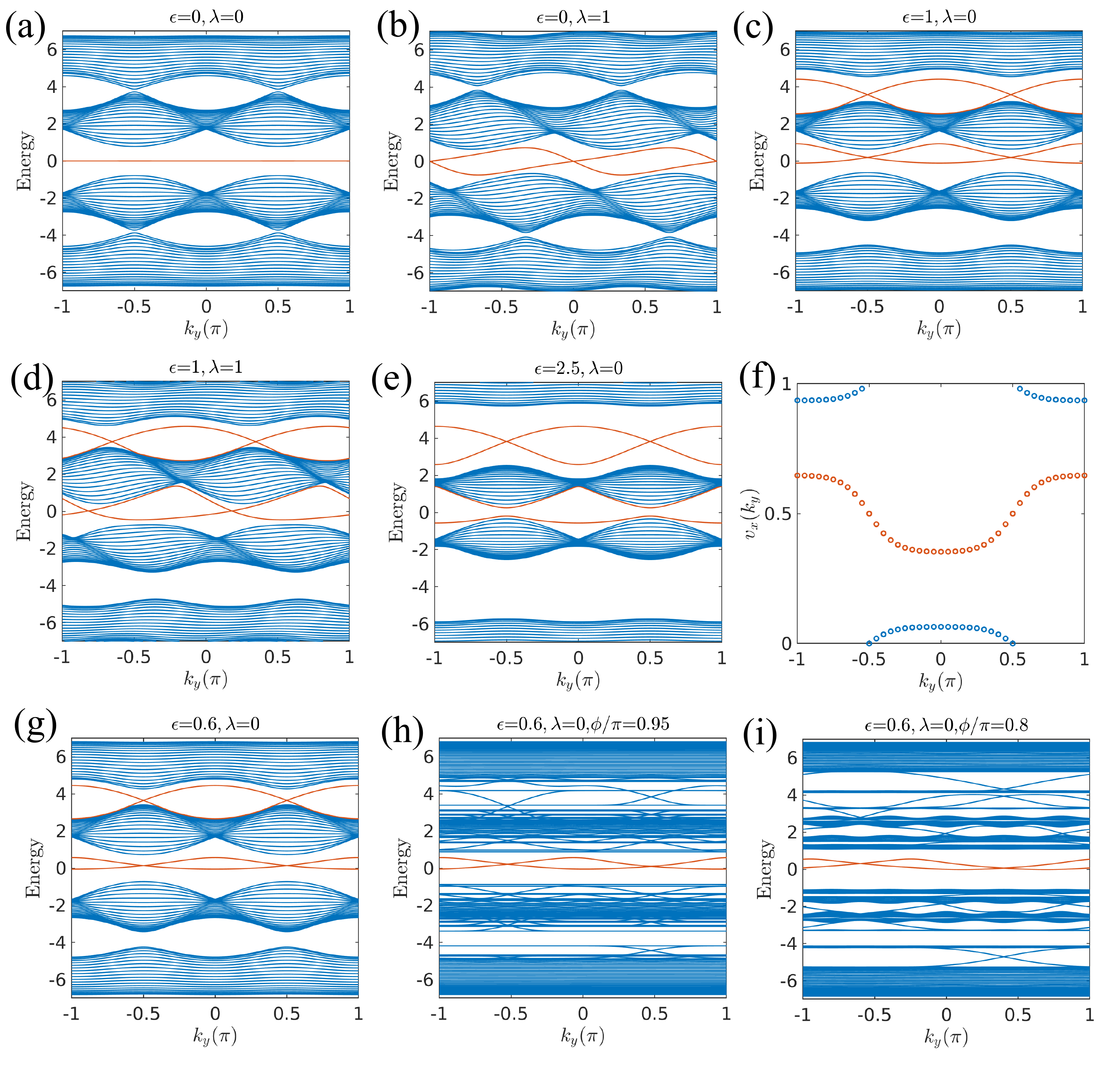}

\caption{(a-e) and (f) Energy spectra of the model specified in Eq\textcolor{black}{.\ \eqref{eq:Zhao}}
on a ribbon along $y$ direction. The red color lines indicate the
twined edge modes. (f) Wannier bands of the lowest energy band (blue)
and the second energy band (red) corresponding to panel (c). (h) and
(i) correspond to (g) but with the flux deviation from $\pi$. The
other parameters are $t_{11}^{x}=t_{22}^{x}=1$, $t_{12}^{x}=t_{21}^{x}=3.5$,
$t_{1}^{y}=2$, and $t_{2}^{y}=1.5$, the same as in Ref. \cite{ChenZ22nc}.
The values of $\epsilon$ and $\lambda$ are labeled on each plot.
\label{fig: Zhaomodel}}
\end{figure}

For a better understanding of our results, we revisit the Klein-bottle
insulator proposed in Ref. \cite{ChenZ22nc}. The Hamiltonian in 2D
reads 

\begin{equation}
H_{0}({\bf k})=\left(\begin{array}{cccc}
\epsilon & [q_{1}^{x}(k_{x})]^{*} & [q_{+}^{y}(k_{y})]^{*} & 0\\
q_{1}^{x}(k_{x}) & \epsilon & 0 & [q_{-}^{y}(k_{y})]^{*}\\
q_{+}^{y}(k_{y}) & 0 & -\epsilon & [q_{2}^{x}(k_{x})]^{*}\\
0 & q_{-}^{y}(k_{y}) & q_{2}^{x}(k_{x}) & -\epsilon
\end{array}\right),\label{eq:Zhao}
\end{equation}
where the parameters are defined as $q_{\ell}^{x}(k_{x})=t_{\ell1}^{x}+t_{\ell2}^{x}e^{ik_{x}}$
with $\ell=1,2$ and $q_{\pm}^{y}(k_{y})=t_{1}^{y}\pm t_{2}^{y}e^{ik_{y}}$.
The staggered on-site potential $\pm\epsilon$ opens a band gap at
the finite-energy Dirac points. The inclusion of the term $H'({\bf k})=\lambda\cos k_{y}\sigma_{1}\tau_{2}+\lambda\sin k_{y}\sigma_{2}\tau_{2}$
breaks time-reversal symmetry. In Fig. \ref{fig: Zhaomodel}, we plot
the spectrum of the model. 

In the limit $\epsilon=0$ and $\lambda=0$, there is an energy gap
close to the zero energy. Then, Dirac points, formed by first and
second (third and fourth) bands, emerge. In this case, the nontrivial
polarization gives zero energy edge modes on a ribbon along $y$ direction
{[}see Fig. \ref{fig: Zhaomodel}(a){]}, similar to the zero-energy
modes in zigzag graphene ribbons \cite{LiCA19jpcm,Ryu02prl}. When
considering the ribbon along $x$ direction, it shows a trivial gap
without edge modes. This anisotropic property is the same as in the
inclined 2D SSH model \cite{LiCA22prr}. 

If we turn on the $\lambda$ term, the flat edge modes become dispersive.
If we turn on the on-site potential $\epsilon$, we find the Dirac
points at finite energy are gapped out. Then, there are two pairs
of twined edge modes: one pair close to zero energy and the other
pair at finite energy. The appearance of twined edge modes can be
understood from the Wannier spectra in Fig. \ref{fig: Zhaomodel}(f).
One branch of Wanner bands exhibits nontrivial winding around $\nu_{x}=\frac{1}{2}$
and the other one exhibits trivial winding around $\nu_{x}=0$ instead.
The total polarization is $p_{x}=\frac{1}{2}$. 

In Fig. \ref{fig: Zhaomodel}(d), we turn on both $\lambda$ and $\epsilon$
terms. It yields the same result as shown in Ref. \cite{ChenZ22nc},
but now we observe two pairs of twined edge modes within a larger
energy window. Tuning the parameter $\epsilon$, this can change the
position of twined edge modes {[}see Fig. \ref{fig: Zhaomodel}(e){]}. 

The twined edge modes also show robustness against flux perturbations.
We consider a case in Fig. \ref{fig: Zhaomodel}(g), where one pair
of twined edge modes are detached from the bulk bands close to zero
energy and the other pair is attached to the bulk continuum at finite
energy. When the flux gradually deviates from $\pi$, we find the
detached twined edge modes persist in the spectra, as shown in Figs.
\ref{fig: Zhaomodel}(h) and \ref{fig: Zhaomodel}(i). The other pair
of edge modes at finite energy start to hybridize with the bulk bands. 

\begin{figure}
\includegraphics[width=1\linewidth]{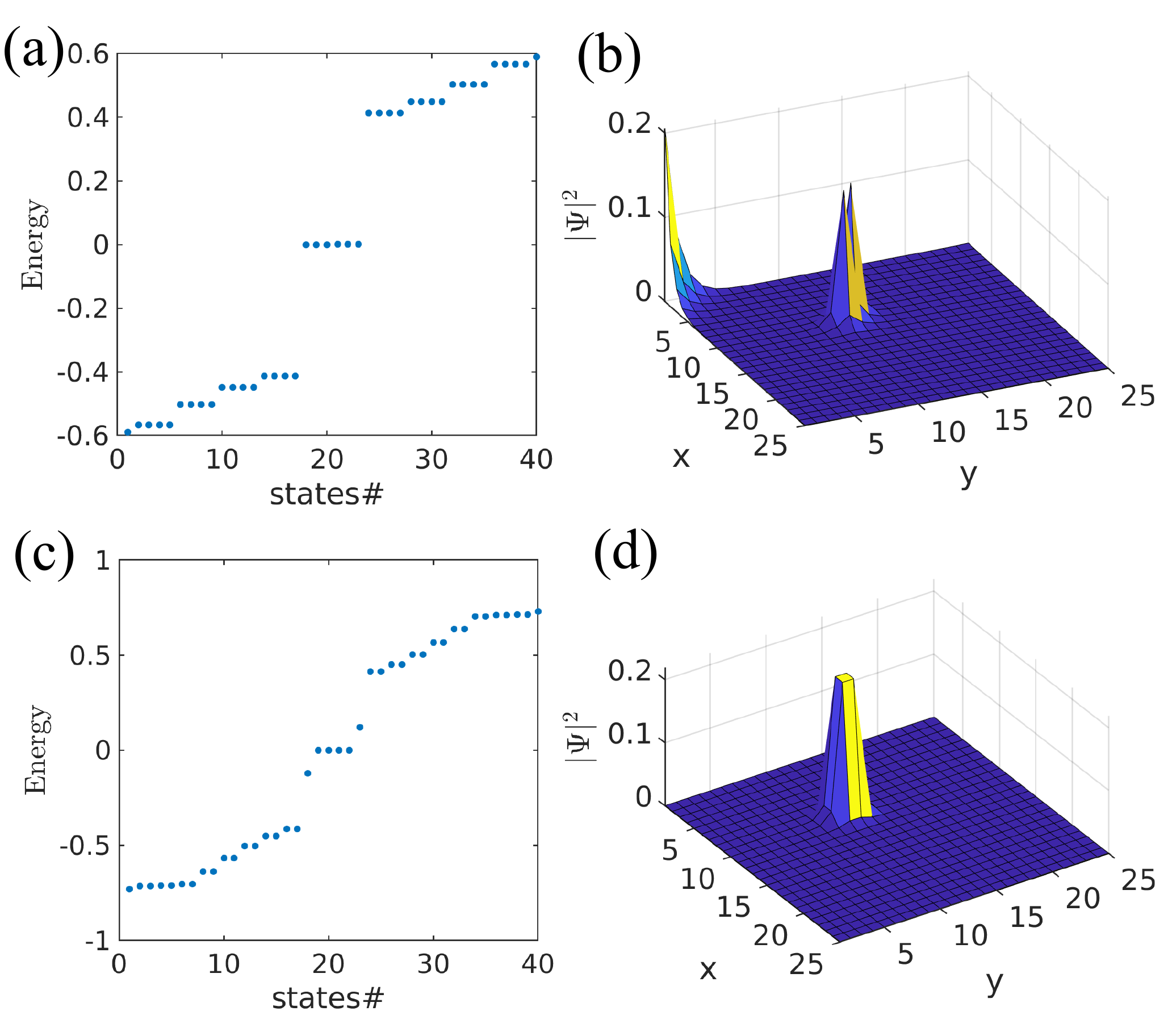}

\caption{(a) Energy spectra for BBH model with a $\pi$-flux defect. There
are totally six zero-energy modes in the gap in the topological nontrivial
phase. The $\pi$-flux defect bounds extra two zero-energy modes.
(b) The wave function distribution corresponds to bound state in (a).
Here $t_{x}=t_{y}=0.5$ for (a) and (b). (c) The same to panel (a)
but the two bound states are away from zero energy. (d) The wave function
distribution corresponds to bound state in (c). Here $t_{x}=0.3,t_{y}=0.6$
for (c) and (d).\label{fig: bound states}}
\end{figure}

\section{Overview of Benalcazar-Bernevig-Hughes model}

For the convenience of comparison with Klein-bottle BBH models presented
in the main text, let us first briefly review the BBH model in 2D
\cite{Benalcazar17Science,BBH17prb}. The tight-binding Hamiltonian
in real space is described as 

\begin{alignat}{1}
H_{0}= & \sum_{\mathbf{R}}\Big[t_{x}(C_{\mathbf{R},1}^{\dagger}C_{\mathbf{R},3}+C_{\mathbf{R},2}^{\dagger}C_{\mathbf{R},4})\nonumber \\
 & +t_{y}(C_{\mathbf{R},1}^{\dagger}C_{\mathbf{R},4}-C_{\mathbf{R},2}^{\dagger}C_{\mathbf{R},3})\nonumber \\
 & +t(C_{\mathbf{R},1}^{\dagger}C_{\mathbf{R}+\hat{x},3}+C_{\mathbf{R},4}^{\dagger}C_{\mathbf{R}+\hat{x},2})\nonumber \\
 & +t(C_{\mathbf{R},1}^{\dagger}C_{\mathbf{R}+\hat{y},4}-C_{\mathbf{R},3}^{\dagger}C_{\mathbf{R}+\hat{y},2})\Big]+\mathrm{H.c}..
\end{alignat}
The corresponding Bloch Hamiltonian in momentum space is 
\begin{alignat}{1}
H_{0}({\bf k}) & =[t_{x}+t\cos k_{x}]\Gamma_{4}+t\sin k_{x}\Gamma_{3}\nonumber \\
 & +[t_{y}+t\cos k_{y}]\Gamma_{2}+t\sin k_{y}\Gamma_{1}.\label{eq:H_q}
\end{alignat}
The Gamma matrices are defined as $\Gamma_{j}\equiv-\tau_{2}\sigma_{j}$,
and $\Gamma_{4}\equiv\tau_{1}\sigma_{0}$. The bulk bands of Eq\textcolor{black}{.\ \eqref{eq:H_q}}
are gapped unless $t_{s}/t=\pm1$ ($s=x,y$). Hence, it is an insulator
at half-filling. The nonspatial symmetries of the BBH model are chiral
symmetry, time-reversal symmetry, and particle-hole symmetry. 

The nontrivial phase of quadrupole insulator is characterized by quantized
quadrupole moments $q_{xy}=\frac{1}{2}$, which induces quantized
corner charge $Q^{\mathrm{corner}}$ and edge polarization $p^{\mathrm{edge}}$
of its equal magnitude $|q_{xy}|=|p_{x}^{\mathrm{edge}}|=|p_{y}^{\mathrm{edge}}|=|Q^{\mathrm{corner}}|$.
The quantization of $q_{xy}$ relies on chiral symmetry \cite{LiCA20prl,YangYB21prb}.
The quadrupole insulators in 2D have boundaries that are stand-alone
1D topological insulators. The nontrivial topological quadrupole phase
is located in the parameter region $|t_{s}/t|<1$ \cite{Benalcazar17Science,BBH17prb}. 

\section{Bound states by $\pi$-flux defects in the original Benalcazar-Bernevig-Hughes
model}

In this section, we demonstrate that a single $\pi$-flux defect in
the original BBH model may trap two bound states. The original BBH
model needs $\pi$ fluxes on all plaquettes of the 2D lattice. A $\pi$-flux
defect means that at a specific plaquette the $\pi$ flux is removed.
Consider a single $\pi$-flux defect in the 2D BBH model lattice.
In the nontrivial quadrupole insulator phase, the $\pi$-flux defect
can induce bound states in the energy gap. As shown in Fig. \ref{fig: bound states}(a),
there are totally six zero-energy modes in the bulk gap: four of them
are corner modes and the extra two are bound states at the $\pi$-flux
defect. These two bound states are degenerate at zero energy. Their
wave function is shown in Fig. \ref{fig: bound states}(b). Another
possibility is that the bound states have finite energy, as shown
in Fig. \ref{fig: bound states}(c). Their wave function localizes
at the position of the $\pi$-flux defect {[}see Fig. \ref{fig: bound states}(d){]}.

\bibliographystyle{apsrev4-1-etal-title}

\end{document}